\documentclass{aa}
\usepackage{graphicx}
\usepackage[varg]{txfonts}
\usepackage{longtable}
\usepackage{lscape}

\begin{document}

\title{Activity time series of old stars from late F to early K. IV. Limits of the correction of radial velocities using chromospheric emission}

\titlerunning{Activity time series of old stars from late F to early K. IV. }

\author{N. Meunier \inst{1}, A.-M. Lagrange \inst{1}, S. Cuzacq \inst{1}
  }
\authorrunning{Meunier et al.}

\institute{
Univ. Grenoble Alpes, CNRS, IPAG, F-38000 Grenoble, France\\
\email{nadege.meunier@univ-grenoble-alpes.fr}
     }

\offprints{N. Meunier}

\date{Received ; Accepted}

\abstract{Inhibition of the convective blueshift in active regions is a major contribution to the radial velocity (RV) variations, at least for solar-like stars. A common technique to correct for this component is to model the RV as a linear function of chromospheric emission, because both are strongly correlated with the coverage by plages.}
{This correction, although efficient, is not perfect: the aim of the present study is to understand the limits of this correction and to improve it. }
{We investigate these questions by analysing a large set of synthetic time series corresponding to old main sequence F6-K4 stars modelled using a consistent set of parameters. We focus here on the analysis of the correlation between time series, in particular between RV (variability due to different processes) and chromospheric emission on different timescales. We also study the temporal variation for each time series. }
{We find that inclination strongly impacts these correlations, as well as the presence of additional signals (in particular granulation and supergranulation). Although RV and $\log R'_{HK}$ are often well correlated, a combination of geometrical effects (butterfly diagrams related to dynamo processes and inclination) and activity level variations over time create an hysteresis pattern during the cycle, which produces a departure from an excellent correlation: for a given activity level, the RV is higher or lower during the ascending phase compared to the descending phase of the cycle depending on inclination, with a reversal for inclinations about 60$^{\circ}$ from pole-on. We find that this hysteresis is also observed for the Sun, as well as for other stars. This property is due to the spatio-temporal distribution of the activity pattern (and therefore to the dynamo processes) and to the difference in projection effects of the RV and chromospheric emission. } 
{ These results allow us to propose a new method which significantly improves the correction 
for long timescales (fraction of the cycle), and could be crucial to improving detection rates 
of planets in the habitable zone around F6-K4 stars.  
}


\maketitle

\section{Introduction}

The detection of low-mass planets using the radial velocity (RV) technique is strongly impacted by the presence of stellar variability. Magnetic activity leads to spurious RV signals around the rotational period as well as on longer timescales (related to cycle variations). We have shown that on long timescales the signal is mostly due to the inhibition of the convective blueshift in plages \cite[][]{meunier10a}, producing not only a strong peak in the periodograms at the cycle period, but also a significant power at higher frequencies (periods of a few hundred days and above): this decreases the detection limit of low-mass planets by one to two orders of magnitudes. A commonly used technique to correct for this contribution relies on the strong correlation between this RV component and the chromospheric emission as measured by the classical $\log R'_{HK}$, since both depend on the filling factor covered by plages. Such an approach has been used in a large number of publications \cite[e.g.][]{boisse09,pont11,dumusque12,robertson14,rajpaul15,lanza16,diaz16,borgniet17}  and allows a significant decrease in the RV jitter to be achieved. This correlation is also used in the fitting challenge organised by X. Dumusque to compare the performance of up-to-date correction techniques \cite[][]{dumusque16,dumusque17}. 

The limitations of this approach are however not well studied.
In the solar case, we have shown that although this approach allows a significant gain, it is not suitable for very low planet masses \cite[][]{meunier13}: for example, at distances of around 1 AU, planets of  1~M$_{\rm Earth}$ could only be reached with very good temporal samplings. It is critical to understand how these limitations depend on the star (spectral type, activity level, ...)  and why a better performance cannot be achieved with this method. The issues are the following: (1) The dependence on $\log R'_{HK}$  can be fitted to a certain extent by a polynomial in time, which is often used  to remove the contribution from possible unknown companions at very long periods that lead to a degeneracy between the $\log R'_{HK}$   and this polynomial. (2) The correction using the $\log R'_{HK}$  time series is not perfect, and a significant amount of power remains in the periodograms, especially at long periods (around the cycle period and below). 

In this paper, we investigate these issues using a large set of simulations covering F6 to K4 old main sequence stars with various activity levels. The generation of the time series is described in \cite{meunier19}, hereafter referred to as Paper I,  which also shows how the $\log R'_{HK}$, representing the chromospheric emission (average level, cycle amplitude) varies with inclination, with results which agree with \cite{shapiro14}. An initial analysis of the RV jitter was made in \cite{meunier19b}, who compared the RV jitter from simulations with observed jitter \cite[][]{saar99,santos00,wright05,isaacson10}. They also showed the good agreement with the RV-R'$_{\rm HK}$ versus spectral type slope observed by \cite{lovis11b} on a large sample of HARPS observations. Finally, they showed that the current status of correction techniques would not allow the detection of Earth-mass planets around stars like the Sun, and that for lower-mass stars a very large number of points was necessary to reach that goal. A significant improvement of the correction method must therefore be made. Physical models are needed to remove the activity contribution to ensure that the residuals are well controlled: the correction to be made is typically one to two orders of magnitude (depending on the activity level). 
We show in the present paper that the time series based on complex and realistic  activity patterns can help us to find clues to develop better models to perform these corrections, which in the end should help to better control the residuals after correction. 

The outline of the paper is the following. In Sect.~2 we briefly recall the model and parameters. In Sect.~3, we study the correlation between RV and $\log R'_{HK}$  time series across our grid of parameters. Subsequently, we present a study of the typical gain which is obtained when performing standard correction using $\log R'_{HK}$  and a polynomial in time. In Sect.~5, we identify a limiting element in the $\log R'_{HK}$  correction and characterize it. We propose a new method to correct for this effect in Sect.~6, which will impact the detectability at long orbital periods. Finally, we conclude in Sect.~7. 

\section{Model and parameters}

Our model, described in detail in \cite{borgniet15} for the Sun and in Paper I for other stars (F6 to K4 and relatively old main sequence stars), provides consistent spots, plages, and network structures for complex activity patterns similar to the Sun.

The time series due to activity is the sum of the contributions due to spots ({\it rvspot1} and {\it rvspot2}, where two laws are used for the spot temperature contrast\footnote{A lower limit defined by the solar contrast in \cite{borgniet15}, $\Delta$Tspot$_1$, and an upper limit law depending on T$_{\rm eff}$ from \cite{berd05}, $\Delta$Tspot$_2$}), plages {\it rvplage} \cite[contrast dependence on spectral type from][]{norris18}, and inhibition of the convective blueshift in plages {\it rvconv}. We also consider the addition of the contribution of oscillation, granulation, and supergranulation  (OGS)  signal, {\it rvogs}, averaged over 1 hour, and an instrumental white noise represented by a Gaussian noise with an amplitude of 0.6 m/s ({\it rvogs}).
Here, $\log R'_{HK}$  time series are produced to be able to relate RV variation to chromospheric emission. We also produced photometric \cite[][]{meunier19d} and astrometric \cite[][]{meunier19f} time series, but the analysis of these time series is outside the scope of the present paper as we focus on the relationship between RV and chromospheric emission. 
The temporal step is one day on average (with random departures of up to 4 hours). The time series have a maximum length of 15 years, and always cover an integer number of cycles. 

The rotation periods, and the cycle periods and amplitudes also depend on the star, and a range of realistic values is considered for each spectral type and activity level. Some parameters which are not constrained are kept to the solar values considered in \cite{borgniet15}, for example meridional circulation  or size distributions. 
We refer to Paper I for more details about the laws used to produce the consistent sets of parameters, where all references and justifications can be found. Of particular interest in this paper is the maximum average latitude at the beginning of cycle $\theta_{\rm max}$: it is not  yet constrained from observations or models, and three values were considered in Paper I, the solar latitude $\theta_{\rm max,\odot}$, $\theta_{\rm max,\odot}$+10$^{\circ}$, and $\theta_{\rm max,\odot}$+20$^{\circ}$ , so that the activity pattern covers different ranges in latitude.

\section{Correlation between RV and $\log R'_{HK}$ }

In the solar case, we showed that inhibition of the convective blueshift was dominating over the spot and plage contributions \cite[][]{meunier10a,borgniet15}. We know from other stars that the correlation is not always strong, as seen for  example from the RV-R'$_{\rm HK}$ slope in \cite{lovis11b}: the slope can be small in some cases \cite[see also][]{meunier17}, and it is important to better understand and quantify how this property varies with spectral type or inclination for example to interpret the observations. It is usually assumed that departures from a perfect correlation are due to the addition of the spot and plage signal for example, but this may not be the only cause. 


We first consider  the correlation\footnote{In the remainder of this paper, the correlation refers to the Pearson correlation computed between two time series. We have also computed Spearman correlations, but they are almost the same, with an rms of the difference between the two over all time series of between 0.01 and 0.05 depending on the configuration, which does not change our conclusions. } between RV and $\log R'_{HK}$, because it can easily be derived from observational data and gives some clues on what to expect from the $\log R'_{HK}$  correction. Typical values are shown in detail in Appendix A.1. 
The correlation between RV and $\log R'_{HK}$  depends on many parameters, but the most influential are spectral type, activity level, and inclination. The departure from a correlation of one explains why using $\log R'_{HK}$  to correct RV times series is not perfect.
In addition, the presence of noise (OGS or instrumental noise) strongly degrades the global correlation. 
It is for example possible to have a RV jitter dominated by {\it rvconv}, although the correlation reaches relatively low values (Appendix A.2). 
This is probably due to the fact that in some cases, the addition of the OGS signal significantly degrades the correlation even though its contribution to the RV jitter is not major.

The global correlation between RV and $\log R'_{HK}$  includes contributions from both short and long timescales. 
The short-term correlations presented in Appendix A.3 are related to the global correlation, but there are a large number of simulations with short-term correlations that are much lower than the global one; these are always lower than 0.75. The inhibition of the convective blueshift is less dominant even for edge-on stars, and there are many configurations where the inhibition of the convective blueshift is not dominant on short timescales.  
In addition, we observe a very strong effect of inclination: 
for pole-on stars, the short-term correlations are much lower than the global ones. This is due to the fact that for low inclinations the rotation modulation is much smaller and therefore the short timescale signal becomes noisier. 

We also note that if the inclination of a given star is well known, it seems possible to relate the global and average short-term correlation. 
The observation of a given short-term correlation at a given time is however not representative of the average short-term correlations, and its sign can even be negative as local correlations cover a wide range of values.
Finally, local correlations computed on solar time series also  show a large dispersion similar to what is observed in our simulations (Appendix A.4). 


\section{Decrease of the gain in RV jitter when correcting using the $\log R'_{HK}$--RV correlation}

To characterize the gain in RV jitter, we use the ratio between the RV jitter before and after correction. The ratio should be higher than one to get an efficient correction, and the higher the ratio, the better the correction. 
We first consider the relationship between gain and global correlation, and analyse how the gain varies for different types of corrections. The first model used to perform the correction is of the form RV=$\alpha$+$\beta$$\log R'_{HK}$. In a second step, we discuss the addition of a polynomial in time, described in the introduction: the model is then of  the form RV=$\alpha$+$\beta$$\log R'_{HK}$ +$\gamma$t+$\delta$t$^2$ \cite[][]{dumusque17}. 

\subsection{Gain while correcting from $\log R'_{HK}$  and polynomial in time}

\begin{figure}
\includegraphics{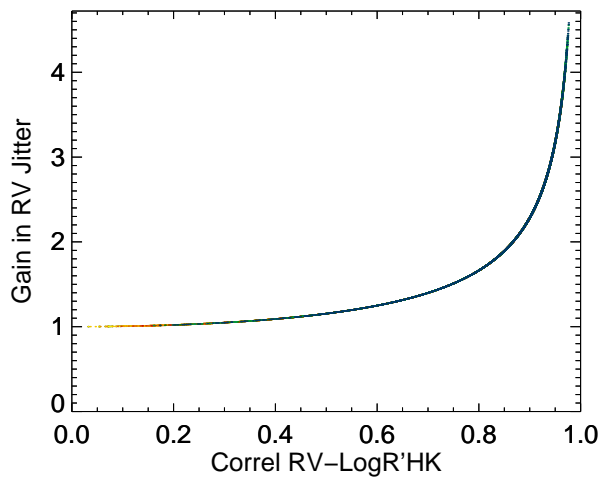}
\caption{
Gain in RV jitter (when correcting for $\log R'_{HK}$  alone) vs. global correlation between RV and $\log R'_{HK}$  (including {\it rvogs} and {\it rvinst}). 
}
\label{gain}
\end{figure}

\begin{figure}
\includegraphics{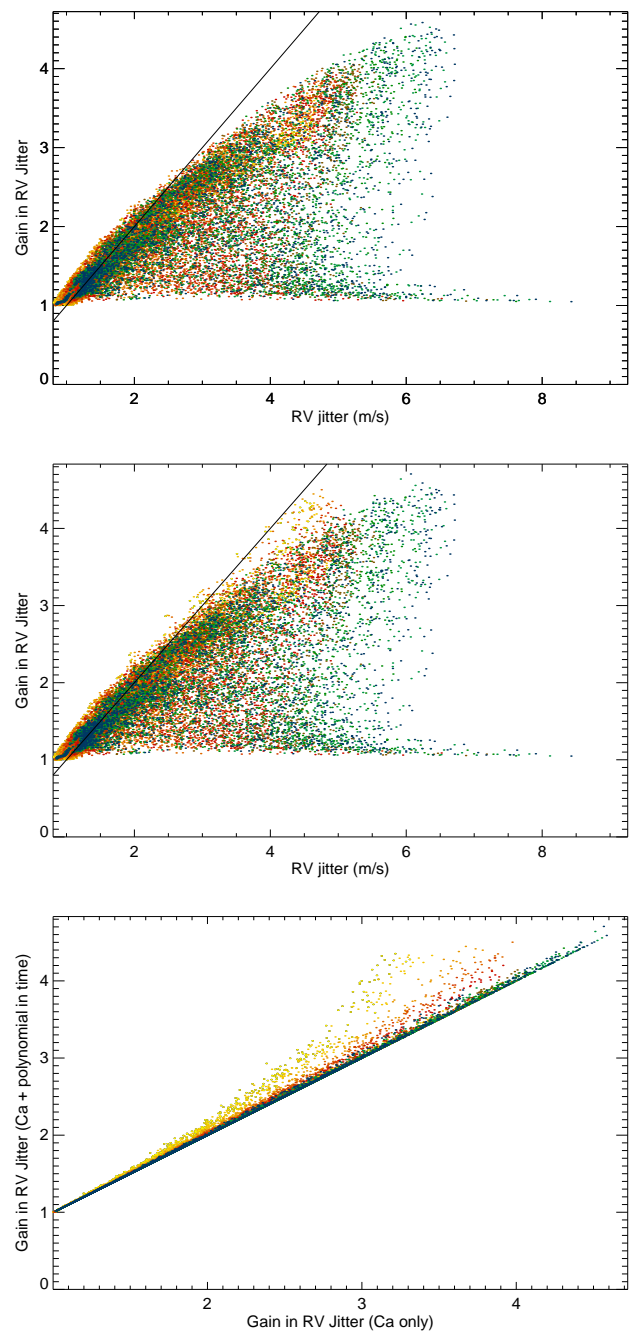}
\caption{
        Gain in RV jitter in various conditions.
{\it Upper panel:} Gain vs. RV jitter (for simulations including {\it rvogs} and {\it rvinst}, and for $\Delta$Tspot2), for a correction with $\log R'_{HK}$  alone. The straight black line has a slope of 1, indicating points where the correction allows a residual RV jitter of 1 m/s to be reached. 
 The colour code corresponds to inclination, from pole-on (i=0$^{\circ}$, yellow) to edge-on (i=90$^{\circ}$, blue), with light and dark orange corresponding to 20$^{\circ}$ and 30$^{\circ}$, light and dark red to 40$^{\circ}$ and 50$^{\circ}$, brown to  60$^{\circ}$, and light and dark green to 70$^{\circ}$ and 80$^{\circ}$.
{\it Middle panel:} Same for a correction with $\log R'_{HK}$  and a second-degree polynomial in time. 
{\it Lower panel:} Gain with $\log R'_{HK}$  and time correction vs. gain with $\log R'_{HK}$  correction only. 
         Only one point out of five is shown for clarity. 
}
\label{gain_rv}
\end{figure}

It is interesting to see whether it is easier to correct for activity using the $\log R'_{HK}$--RV correlation when the correlation between RV and $\log R'_{HK}$  is high. Figure~\ref{gain} shows the gain defined at the beginning of Sect.~4 versus the $\log R'_{HK}$--RV correlation (from Sect.~3) and  indeed shows a perfect relationship between the two approaches. When the correlation departs from one, the gain also decreases.

The gain as a function of the original RV jitter (before correction) is shown in Fig.~\ref{gain_rv}. 
The upper panel corresponds to a simple correction using a linear relation between RV and $\log R'_{HK}$  and we see a complex behaviour with a wide spread: a large number of simulations correspond to low to medium RV jitter and a gain close to the solid line (gain providing a RV jitter of below 1 m/s after correction). For RV jitters in the range 1-2.5 m/s, 37 \% can be corrected to a level below 1 m/s. However, for medium to high RV jitters, the gain can be very different for a given RV jitter: for example, for a jitter of 4~m/s, the gain can vary between 1 (no improvement brought by the correction) and 3.5. The few simulations with the highest RV jitter are very poorly corrected (gain close to 1). Simulations with a RV jitter originally above 2.5 m/s never reach RV jitter below 1 m/s after correction. Finally, the shape of the upper envelope shows that the gain does not vary linearly with RV jitter. 



The second panel shows the same plot, but this time the RV signal is corrected for both $\log R'_{HK}$  and a second degree polynomial in time (which is often done to correct for long-term trends due to companions). The general behaviour is similar, although it is possible in some cases to obtain higher gains, as illustrated on the last panel,  which shows the gain with $\log R'_{HK}$  and polynomial in time correction vs. gain with $\log R'_{HK}$  correction only. This is puzzling, because in the present case, the polynomial in time is not a good model of the time series, since it is not present in our simulations (no added companions): when correcting using this polynomial in time in addition to the linear correlation with $\log R'_{HK}$, we use a  model which is not physical. 




\subsection{Discussion on the polynomial function fitted to the RV time series}

\begin{figure}
\includegraphics{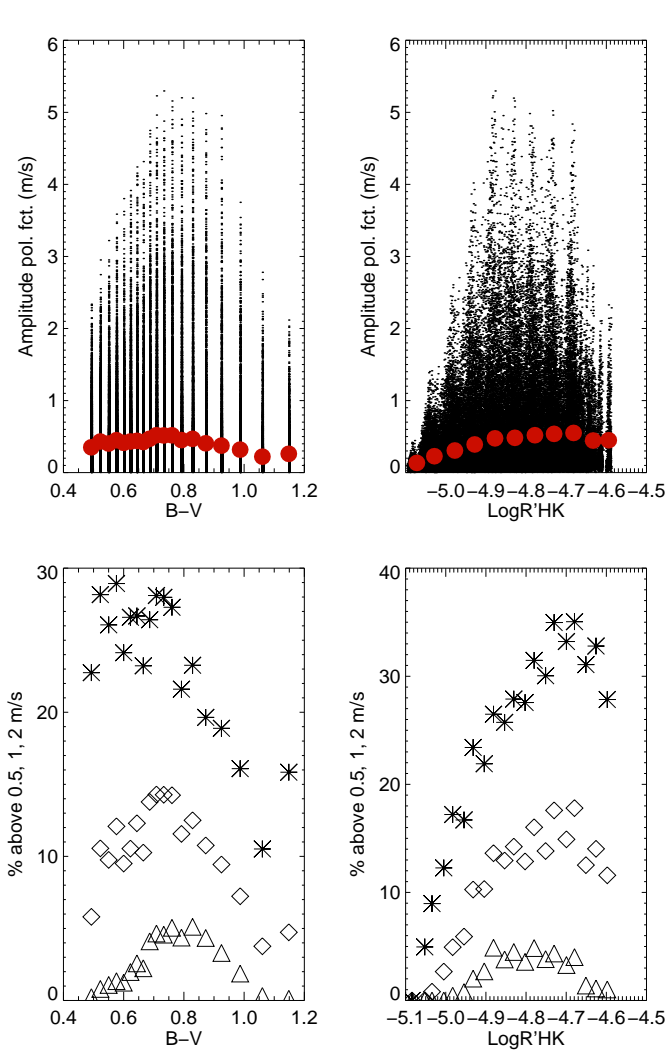}
\caption{
{\it Upper panels:} Amplitude of the polynomial in time vs. B-V (left) and $\log R'_{HK}$  (right). The black dots represent each individual time series, while the red points correspond to a binning in B-V and $\log R'_{HK}$  respectively. 
{\it Lower panels:} Percentage of polynomial amplitudes higher than 0.5~m/s (stars), 1~m/s (diamonds), and 2~m/s (triangles) vs. B-V and $\log R'_{HK}$  respectively.
}
\label{pol}
\end{figure}

\begin{figure}
\includegraphics{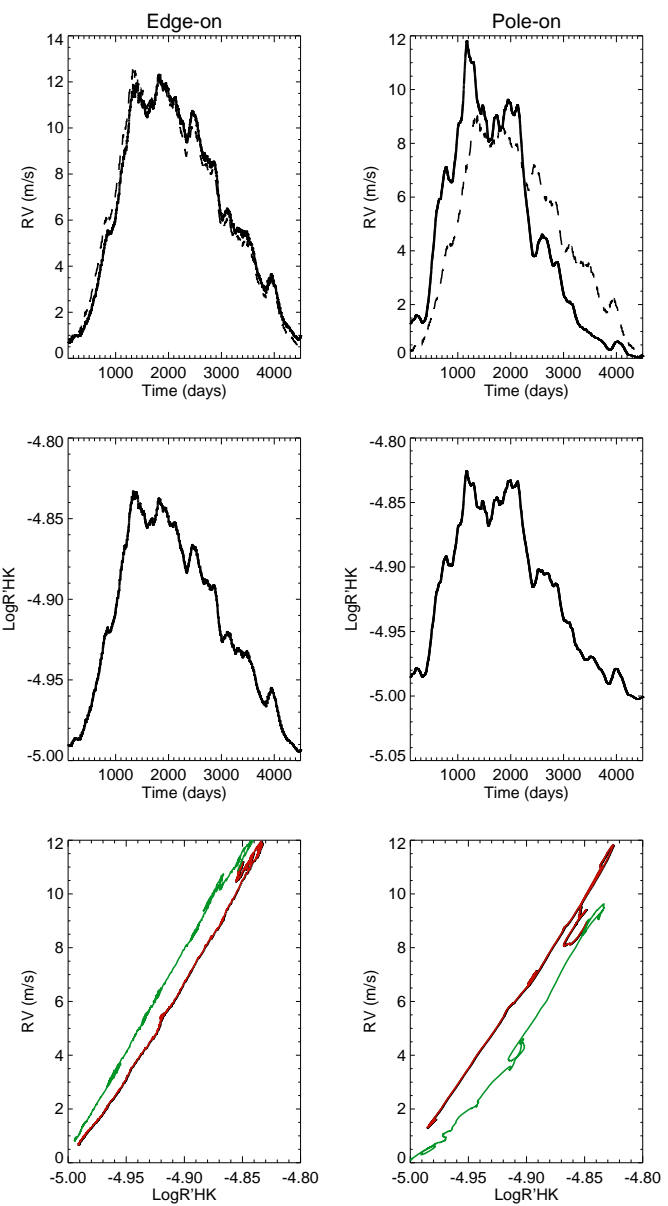}
\caption{
Example of smoothed time series for a moderately active G2 star seen edge-on (left) and pole-on (right). {\it First line:} RV vs. time (solid line), and RV fitted from the $\log R'_{HK}$  variations (dashed line). {\it Second line:} $\log R'_{HK}$  vs. time. {\it First line:} RV vs. $\log R'_{HK}$, during the ascending phase of the cycle (red) and during the descending phase (green). 
}
\label{ex2}
\end{figure}

Given the difference in gain observed for certain simulations when correcting for $\log R'_{HK}$  alone and when also introducing a polynomial in time, it is important to quantify this effect and understand its origin, especially since it is widely used. Furthermore, the simulations do not include companions and therefore the coefficients of the polynomial should be equal to zero (i.e. the polynomial vs. time should be flat); if the coefficients of the polynomial in time are different from zero, this means that we fit the time series with a model which is not physically realistic.

For each simulation, we therefore corrected for $\log R'_{HK}$  and this polynomial in time as above. We then computed the amplitude of the fitted polynomial  A$_{\rm pol}$, defined as the maximum minus the minimum of the polynomial over the time range.  Figure~\ref{pol} shows this amplitude as a function of B-V and $\log R'_{HK}$  (upper panels). Although the average is low, there are simulations with high values of A$_{\rm pol}$, up to a few m/s. The percentage of simulations with amplitudes above 0.5, 1, and 2~m/s is shown in the lower panels. 
For the higher masses and more active (from their average activity level) stars in the lower panels of Fig.~\ref{pol}, the percentage can be relatively high, for example more than 30\% for the 0.5 m/s threshold. The percentage can reach a few percent for the 2 m/s threshold. 

All simulations with high polynomial amplitudes correspond to long cycles, that is, simulations for which only one cycle is simulated, and the highest amplitudes correspond to high cycle amplitudes. For such simulations, the polynomial in time can mimic activity (at least to a certain point, since the shape of the cycles is not exactly polynomial). Since some fits are better with this polynomial than without, the RV variations contain a component which is closer to such a polynomial (as far as a single cycle is concerned) than the $\log R'_{HK}$  variability: a polynomial fit is then  ad hoc and the time series are not well constrained in these cases. We explain this result in the following section.

\section{Hysteresis between RV and $\log R'_{HK}$  time series}

In this section, we explain the previous results by the presence of an hysteresis pattern between the RV and chromospheric emission variabilities. We quantify the amplitude of this pattern as a function of the stellar parameters in our simulations. Finally, we show that it is also observed for the Sun and other stars.

\subsection{Why does the polynomial in time improve the correction?}

\begin{figure*}
\includegraphics{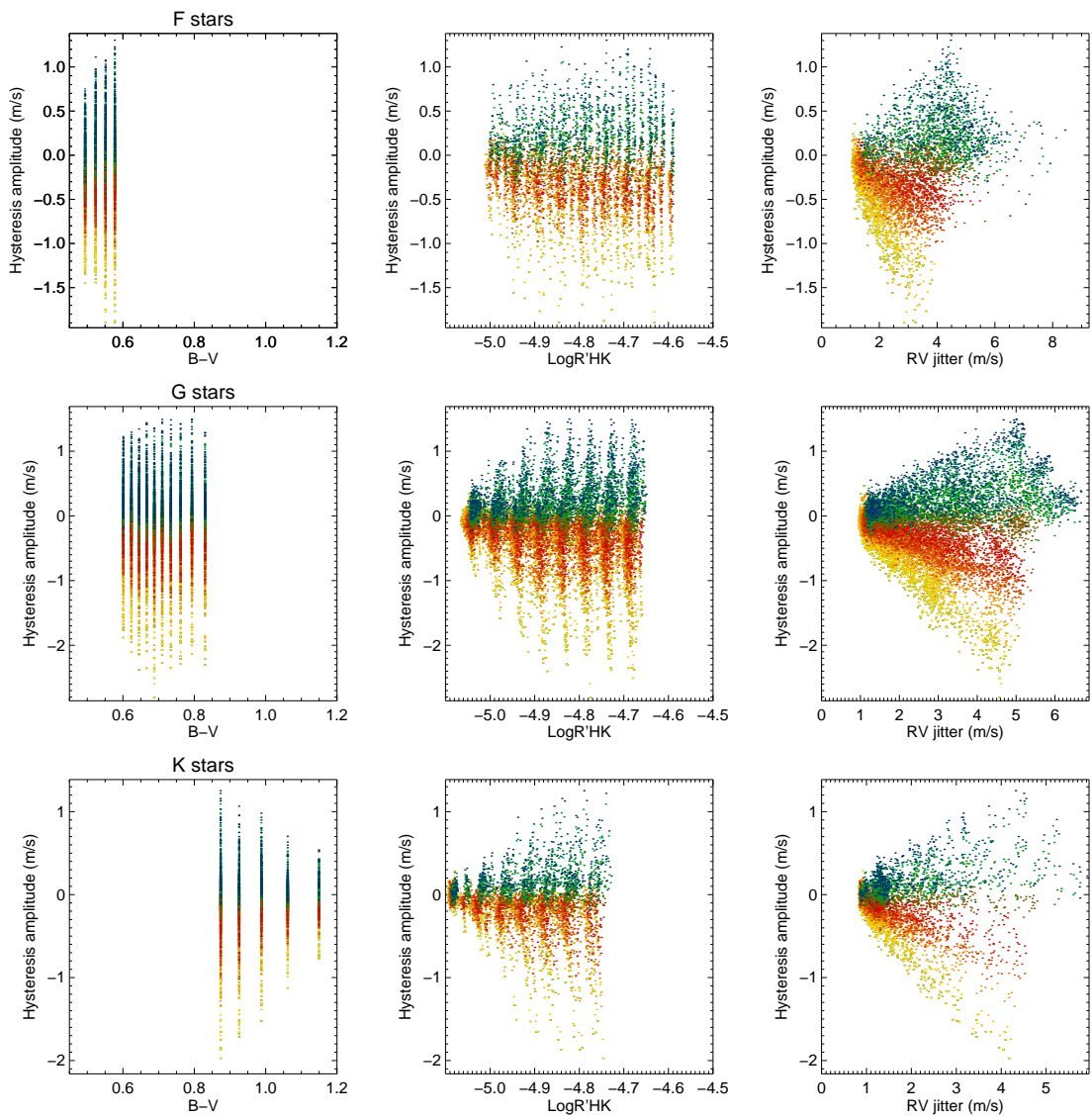}
\caption{
        {\it Upper panels:} Hysteresis amplitude $C_{\rm hyst}$ vs. B-V (left), vs. $\log R'_{HK}$  (middle), and vs. RV jitter (right), for F stars. The colour code is similar to that in Fig.~\ref{gain_rv}. 
{\it Middle panels:} Same for G stars.
{\it Lower panels:} Same for K stars.
         Only one point out of five is shown for clarity. 
}
\label{hyst}
\end{figure*}

\begin{figure}
\includegraphics{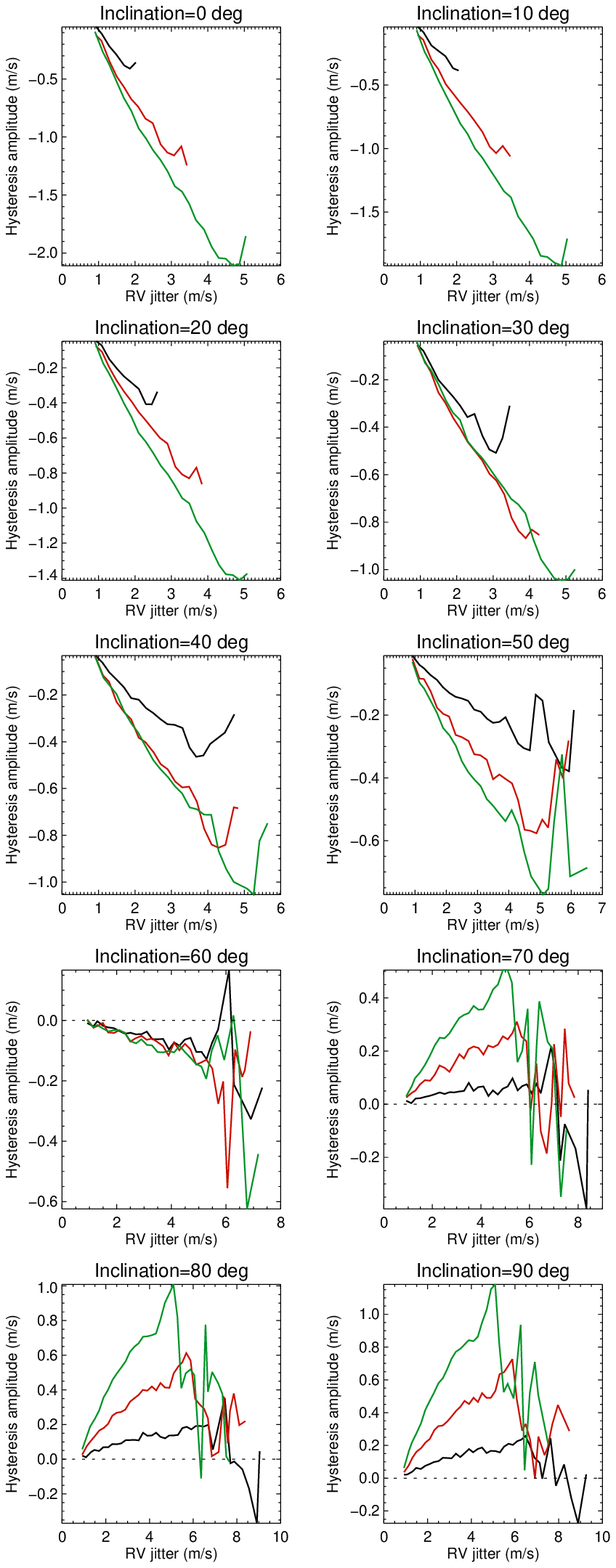}
\caption{
Average hysteresis vs. RV jitter for three values of $\theta_{\rm max}$: solar (black), solar +10$^{\circ}$ (red), and solar +20$^{\circ}$ (green). Each plot corresponds to a different inclination, from pole-on (0$^{\circ}$, upper left panel) to edge-on (90$^{\circ}$, lower right panel).
All plots are on the same scale in m/s.
}
\label{hyst2}
\end{figure}

We first show an example to illustrate the potential impact of the polynomial fit. 
All plots are for $\Delta$Tspot$_2$, and include activity, OGS (smoothed over 1 hour), and instrumental white noise.
Figure~\ref{ex2} shows an example of a time series for a G2 star with a medium activity level: all points (full cycle) and after smoothing for clarity. The upper panel shows the difference between the RV we want to correct and the RV that would be obtained if a linear relationship between RV and $\log R'_{HK}$   were used. The lower panel shows RV versus $\log R'_{HK}$: although there is indeed a very good global correlation between the two (around 0.9  in this example, hence the usual correction method), there is a departure from a strict correlation, with an asymmetry between the ascending phase of the cycle and the descending phase, particularly visible for the pole-on (right panels) plot. For the star seen edge-on, for a given $\log R'_{HK}$  level, the RV is  higher during the descending phase of the cycle compared to the ascending phase. It is the opposite for the same star seen pole-on. 
This is a major effect because we observe differences higher than 1~m/s in these examples, which should lead to an important power in the periodograms of the time series.

There is therefore a kind of hysteresis (we use this denomination in the following), with the two phases of the cycle behaving in a different manner. This effect is not taken into account when  modelling RV as a linear function of $\log R'_{HK}$, but in some cases can be partially taken into account using the polynomial in time. 
There also might be  a systematic effect of inclination. 
We characterize this hysteresis in the following section for all simulations. 


\subsection{Characterisation of the hysteresis across the grid of parameters}

In order to characterise the hysteresis, we define the  criterium $C_{\rm hyst}$. For a given simulation, we consider one cycle. We estimate the position of cycle maximum to separate the ascending phase and the descending phase. We then consider a series of 30 equally spaced $\log R'_{HK}$  levels (corresponding to the range covered by each time series). The RV is averaged over each $\log R'_{HK}$ bin during the ascending phase (RV$_{\rm asc}$) and during the descending phase  (RV$_{\rm desc}$). We then compute RV$_{\rm desc}$-RV$_{\rm asc}$ and average it over the 30 levels to produce $C_{\rm hyst}$. 


Figure~\ref{hyst} shows $C_{\rm hyst}$ versus B-V, $\log R'_{HK}$, and RV jitter. A first  striking result is that there is indeed a systematic effect of inclination, going from a negative (pole-on) to a positive (edge-on) value. The reversal happens around 60$^{\circ}$ from pole-on. $C_{\rm hyst}$ is strongly related to the average activity level and RV jitter (i.e. variability level). Because of this, $C_{\rm hyst}$ naturally decreases towards lower  masses.

The value of $\theta_{\rm max}$  (maximum average latitude at the begining of the cycle) also has a strong effect on the hysteresis.  Figure~\ref{hyst2} shows binned $C_{\rm hyst}$ vs. the RV jitter, for our three values of $\theta_{\rm max}$. In addition to an almost linear relationship with the RV jitter in most cases, $\theta_{\rm max}$ strongly impacts $C_{\rm hyst}$ as well: higher $\theta_{\rm max}$ means a wider range covered by $C_{\rm hyst}$. Finally, the amplitudes are lower for edge-on configurations compared to pole-on configurations. 


In summary, $C_{\rm hyst}$ is strongly related to the activity level, inclinations, and  $\theta_{\rm max}$. 
We attribute this dependence to the conjunction of two facts: structures are not at the same position in latitude on the disk during the cycle, and projection effects are different for RV and $\log R'_{HK}$. This is detailed and discussed in Sect.~6, where a new correction method is proposed.

\subsection{Is the hysteresis present in observations?}

We show that the hysteresis is also observed for the Sun, based on the analysis of two cycle-long time series. We then find that it is also present for other stars.

\subsubsection{The solar case}

\begin{figure}
\includegraphics{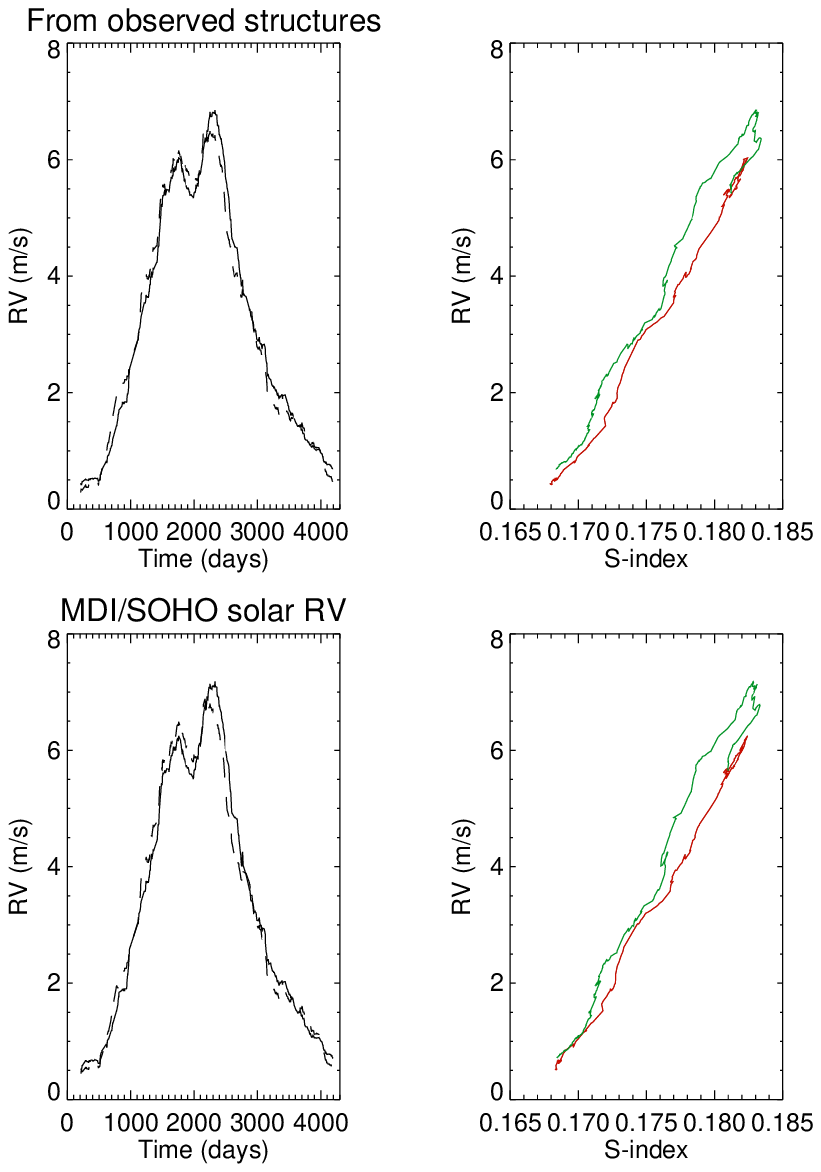}
\caption{
        Relative velocity (solid line), RV deduced from the linear relation with Sacramento Peak S-index (dashed line) vs. time (left panels), and RV vs. S-index (right panels, in red for ascending phase and in green for descending phase). 
{\it Upper panels:} RVs reconstructed from observed structures \cite[][]{meunier10a}. 
{\it Lower panels:} RVs  reconstructed from MDI/SOHO Dopplergrams \cite[][]{meunier10}. 
}
\label{hyst_sol}
\end{figure}

Our simulations provide evidence for the presence of an hysteresis pattern between the long-term
RV and $\log R'_{HK}$  variations over the cycle. We now examine two solar RV time series to confirm this property with observations. 

We first consider the RV reconstruction over a complete cycle made by \cite{meunier10a}. This reconstruction was based on observed solar structures (spots, plages), and a model was used to build the integrated RV from the estimated RV for each structure. After selecting days for which an observation of the chromospheric emission was available (S-index from the Sacramento Peak Observatory), we plot the hysteresis pattern in Fig.~\ref{hyst_sol} (upper panels). The amplitude of the hysteresis is of the order of 0.5 m/s, which is very similar to our simulation for edge-on configurations, both in sign and amplitude. 

We also computed the hysteresis pattern from the solar RV time series reconstructed from SOHO/MDI Dopplergrams \cite[][]{Smdi95} by \cite{meunier10}. In this case, the RV due to active regions was reconstructed by integrating the Doppler velocities over the disk; we did not compute the RV using the model for the RV associated to each structure. The chromospheric emission is also from the Sacramento Peak Observatory. The hysteresis shown in Fig.~\ref{hyst_sol} (lower panels) is very similar to the previous case, which shows that the model is very good to that level of detail. 

These two solar series are, to our knowledge, the only ones available to compare with our simulation. The ongoing solar programs with HARPS-N \cite[][]{dumusque15,collier19,milbourne19} and HARPS produce high-cadence observations of the Sun in stellar conditions, and will be suitable for such an analysis, but the temporal coverage so far is still insufficient to allow such a study (only the end of the descending phase of the solar cycle is available, with a solar minimum in 2019). The reconstruction of the solar RV from structures derived from HMI observations by \cite{milbourne19} is similar in length.

\subsubsection{Stellar observations}

\begin{table}
\caption{Star sample and hysteresis}
\label{tab_star}
\begin{center}
\renewcommand{\footnoterule}{}  
\begin{tabular}{llll}
\hline
Star        & Spectral & Number    & Hysteresis \\ \hline
            &   type   & of points &   \\ \hline
HD21693     & G9 IV-V  & 212 & E\\
HD7199      & K1 IV    & 112 & O\\
HD1461      & G3 V     & 461 & M\\
HD20003     & G8 V     & 183 & P\\
HD207129    & G2 V     & 362 & P\\
HD38858     & G2 V     & 213 & E\\
HD71835     & G9 V     & 109 & O\\
HD82516     & K2 V     & 89 & E\\
HD95456     & F8 V     & 244 & P\\
HD10180     & G1 V     & 327 & M\\
HD13808     & K2 V     & 244 & E\\
\hline
\end{tabular}
\end{center}
\tablefoot{The spectral type is from the CDS (https://simbad.u-strasbg.fr/simbad/). "E" indicates an hysteresis pattern similar to our edge-on configurations, "P" similar to our pole-on configurations, "M" a mixed pattern, and "O" no hysteresis. }
\end{table}

\begin{figure*}
\includegraphics{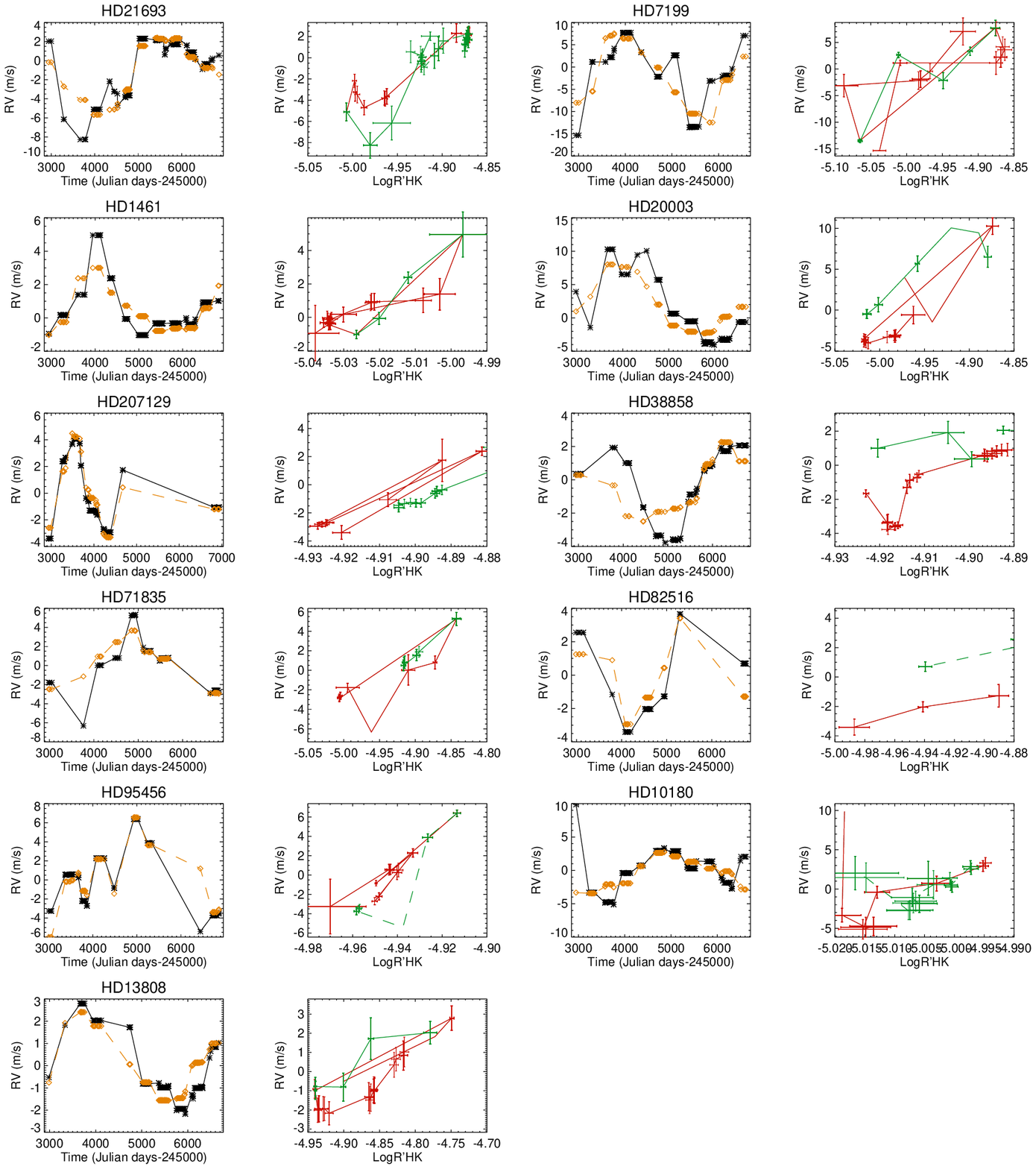}
\caption{
Relative velocity vs. time (stars linked by solid black lines), and RV from linear dependence on $\log R'_{HK}$ (diamonds linked by orange dashed line) for each star. The second plot for each star shows the hysteresis pattern, with the ascending phase in red and descending phase in green. 
}
\label{hyst_star}
\end{figure*}

It is more difficult to check whether such an hysteresis is observed on other stars because of the usually poor temporal sampling. Surveys such as the Mount Wilson  survey provide large samples of stars with a good cycle coverage \cite[e.g.][]{baliunas95}, but they are not associated to simultaneous RV measurements. We have however identified eleven stars observed with HARPS for which a complete cycle with a large number of points is available (Table~\ref{tab_star}). Unlike in our  simulations, the cycle is not necessarily covered from one minimum to the next.   We binned the data over one year, and plot the corresponding hysteresis pattern.
The results are shown in Fig.~\ref{hyst_star}. We observe both types (edge-on or pole-on sign) of hysteresis in seven stars, and four stars show either a more complex mixed pattern (two stars) or no hysteresis (two stars). This is summarised in Table.~\ref{tab_star}. The presence of stars with no hysteresis is expected given the reversal observed in Fig.~\ref{hyst}. A mixed pattern could be due to a spatio-temporal distribution of the structures that is more complex than the solar one. We therefore find evidence suggesting that the same hysteresis is present in other stars as well. 


\section{Towards a better correction of the long-term RV variability}

In this section, we explain the origin of the hysteresis pattern. We then use these results to propose a new correction method using a better model for the relationship between RV and $\log R'_{HK}$, and  illustrate its performance on a subset of G2 star simulations.

\subsection{Explanation of the inclination-dependent hysteresis pattern}

\begin{figure}
\includegraphics{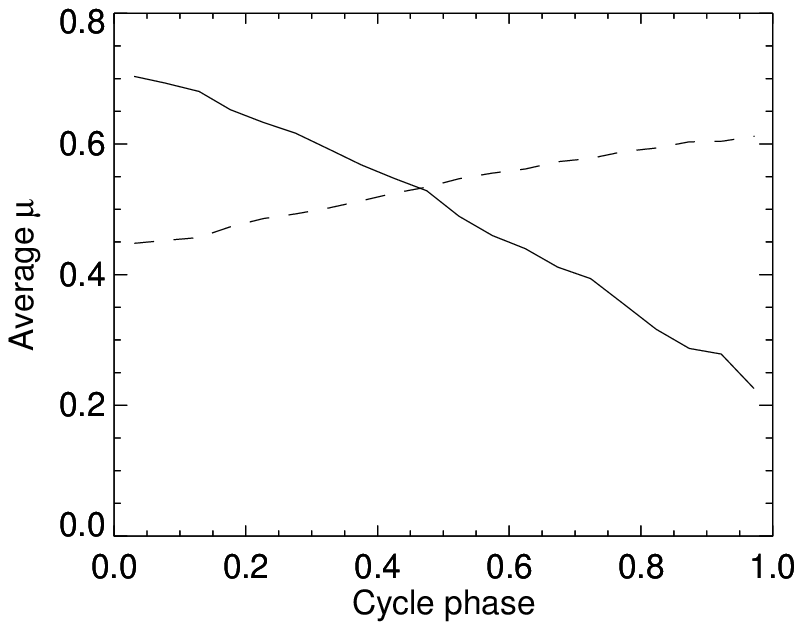}
\caption{
Average $\bar{\mu}$ vs. cycle phase for pole-on (solid line) and edge-on (dashed line) configurations for simulation shown in Fig.~\ref{ex2}. 
}
\label{mumoy}
\end{figure}

\begin{figure}
\includegraphics{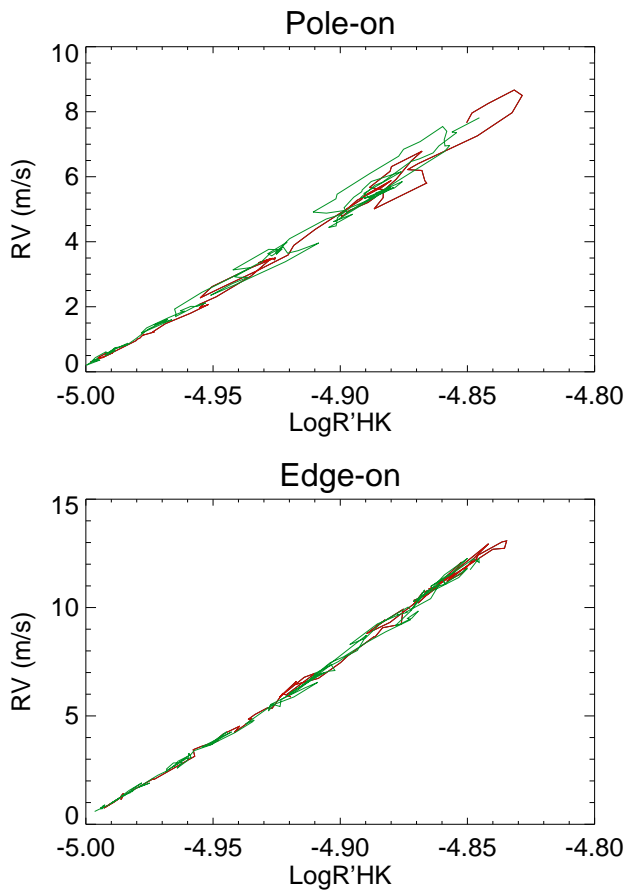}
\caption{
        Relative velocity vs. $\log R'_{HK}$  for simulation similar to Fig.~\ref{ex2}, seen pole-on (upper panel) and edge-on (lower panel), but with constant latitude over time.
}
\label{latcst}
\end{figure}

\begin{figure}[ht]
\includegraphics{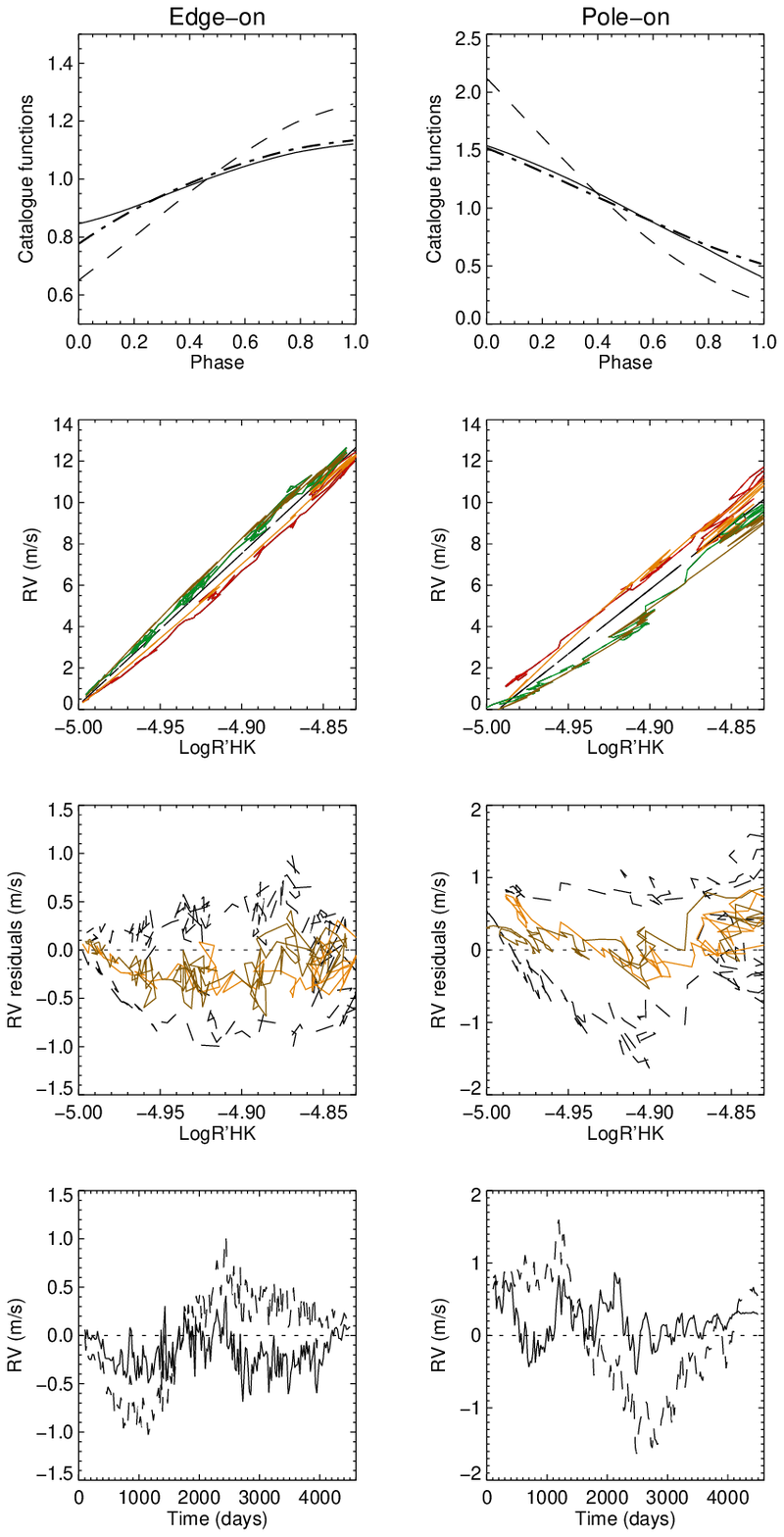}
\caption{
{\it First panel:} Example of correcting function from catalogue (see Sect.~6.4) vs. phase for edge-on configuration (left) and pole-on configuration (right): chromospheric emission (solid line), RV (dashed line), and ratio between RV and chromospheric (thick dotted-dashed line), all normalized to 1 for clarity.  
{\it Second panel:} RV vs. chromospheric emission for the simulation shown in Fig.~\ref{ex2}, colour-coded according to the cycle phase: red for the ascending phase, green for the descending phase. The orange and brown curves correspond to the same phases respectively but for the model derived from the catalogue functions. The black dashed line is the linear fit between RV and $\log R'_{HK}$.
{\it Third panel:} RV residual after standard correction (black dashed line) and after the new correction (orange and brown solid lines as in the second panel). 
{\it Fourth panel:} RV residual vs. time after standard correction (dashed line) and after the new correction (solid line). 
}
\label{catal}
\end{figure}

Our interpretation of this hysteresis pattern is that although RV and chromospheric emission 
are roughly correlated in the long term due to the major contribution of {\it rvconv}, the signal produced by a given active region strongly depends on its position on the disk, while the 
projection effects are different for both variables: the chromospheric emission is essentially linear 
vs. the projected area of the structure. In RV, the dependence on $\mu$ (cosine of the angle between the normal to 
the surface and the line-of-sight) is more complex, as it includes an additional projection effect of the velocity field ($\mu$), 
a contrast dependence on $\mu,$ and the center-to-limb darkening function (see Paper I). 

As a consequence, if the average $\mu$ (hereafter $\bar{\mu}$) changes in time (in particular due to a butterfly diagram 
pattern), this must introduce a departure from the linear relationship between RV and chromospheric 
emission. This is the case when considering different stellar inclinations. 
For a solar-like butterfly diagram such as in our simulations (i.e. toward the equator), 
if the star is seen pole-on, $\bar{\mu}$ decreases during the cycle because the structures appear at 
lower latitudes, as shown in Fig.~\ref{mumoy}, and $\bar{\mu}$ therefore also covers a wide range: RV decreases faster than the chromospheric emission, so 
for a given activity level, the RV signal is higher during the ascending phase of the cycle. 
This is the opposite for a star seen edge-on because during the cycle, $\bar{\mu}$ increases and covers a small range of values: 
for a given activity level, the RV signal is lower  during the ascending phase of the cycle.  
This  behaviour should be responsible for the reversal seen in Fig.~\ref{hyst} and is related 
to the average position of the structures, which is due to the presence of a butterfly diagram pattern, and therefore to dynamo processes. 
It could also explain why $\theta_{\rm max}$ has a strong effect on the hysteresis pattern. 

Furthermore, a simulation similar to the one shown in Fig.~\ref{ex2}, but with a constant latitude over time 
(flat butterfly diagram), exhibits no hysteresis, as illustrated in Fig.~\ref{latcst}. This shows 
that the hysteresis is strongly related to the spatio-temporal distribution of the structures. We also 
note from this figure that the relationship between RV and $\log R'_{HK}$, although showing little 
dispersion, is not strictly linear and some dispersion is still present.

\subsection{Construction of a reference catalogue}

If the butterfly diagram of a star were known, the difference in trend between RV and chromospheric emission could then be  modelled and corrected.
In practice however, it is not known, but we can build a large number of functions corresponding to different configurations (different latitude ranges for example), which can be generated to build a reference catalogue describing the different possibilities for the ratio between the RV and $\log R'_{HK}$ behaviour over time. This must then be modulated by the activity level over time when applied to a given time series. The function leading to the lowest residuals after correction is then selected. We built these functions for the following parameters:
\begin{itemize}
        \item{Stellar inclination: we consider 91 values between 0$^{\circ}$ and 90$^{\circ}$ with a step of 1$^{\circ}$.}
\item{Average latitude at the beginning of the cycle $\theta_{\rm max}$: we consider values between 15$^{\circ}$ and 59$^{\circ}$. 
We recall that in our simulations, we only considered input values of 22$^{\circ}$, 32$^{\circ}$, and 42$^{\circ}$  (the actual average latitudes are usually higher, in particular due to meridional circulation). Here we generate the functions for a  wider range of 
parameters since $\theta_{\rm max}$ is not constrained for a particular observation. }
\item{Width of the butterfly diagram in latitude, $\Gamma$: we considered values from 2 to 20$^{\circ}$, with a step of 2$^{\circ}$. We recall that all 
simulations were made with an input value of 6$^{\circ}$ (the actual values are usually higher due to diffusion and meridional circulation). We use a Gaussian distribution around the average latitude, with a cut at $\pm$20$^{\circ}$.}
\item{Average latitude at the end of the cycle $\theta_{\rm min}$: this is kept constant (9$^{\circ}$, corresponding to the input value in our simulations) in this first analysis. }
\end{itemize}

For each of these parameter sets, we computed the butterfly diagram pattern in latitude for 100 phases during a cycle. We attributed to each pixel of the stellar disk the 
projection effects corresponding to RV and chromospheric emission respectively. 
The center-to-limb darkening is similar to the function used in our simulation in Paper I \cite[][]{claret03}. The plage contrast is an average as a function of $\mu$ of the function used in Paper I. 
For a given inclination, adding the pixel contributions over the whole disk at each phase produces two time series, hereafter RV$_{\rm cat}$ and Ca$_{\rm cat}$.
These functions do not include any structures such as spots or plages: they only describe the relative variability of the two variables (RV and chromospheric emission),
for a constant activity level, due to the position of the structure. For that reason, we also attribute to each pixel a factor describing the fact that 
a structure rotating in longitude would spend more time at a position close to the limb compared to disk center due to projection effects (for a given rotation rate). 
This leads to a catalogue of 20930 functions. 
Figure~\ref{catal} shows an example of such a function (upper panels).

\subsection{Fitting the time series using the reference catalogue}

The catalogue is then used as follows for a given time series covering a stellar cycle: 
\begin{itemize}
	\item{We first bin the time series over the rotation period, because the catalogue is used to improve the correction over  long timescales (we  average the structure positions in longitude). This leads to RV$_{\rm bin}$ and Ca$_{\rm bin}$ vs. t$_{\rm bin}$ in the following. In principle, the same procedure 
could be applied directly to the original time series, but this approach is faster and gives a very good idea of the performance when considering the long-term variations only.}
\item{Each function of the catalogue is then tested: we apply a linear fit to (RV$_{\rm bin}$,Ca$_{\rm bin}$), leading to a RV time series corrected from the linear correlation with Ca$_{\rm bin}$, 
RV$_{\rm corlin}$. Here, RV$_{\rm corlin}$ is then multiplied by RV$_{\rm cat}$/Ca$_{\rm cat}$ (after interpolating on t$_{\rm bin}$) 
                to account for the difference in behaviour with $\mu$ over the cycle of RV and chromospheric emission, and by a constant to minimize the 
residuals, the amplitude in the catalogue being arbitrary. }
\item{Finally, we select the catalogue function providing the lowest rms of the residuals. }
\end{itemize}





When applied to the binned time series corresponding to Fig.~\ref{ex2}, the rms RV before correction is 3.79~m/s (3.65~m/s), after a standard linear correction 0.48~m/s (0.77~m/s), and after the new correction 0.2~m/s (0.27~m/s), for the edge-on and the pole-on configurations respectively. There is therefore a gain of 2.40 (2.85) between the standard correction and the new correction. 


\begin{figure}
\includegraphics{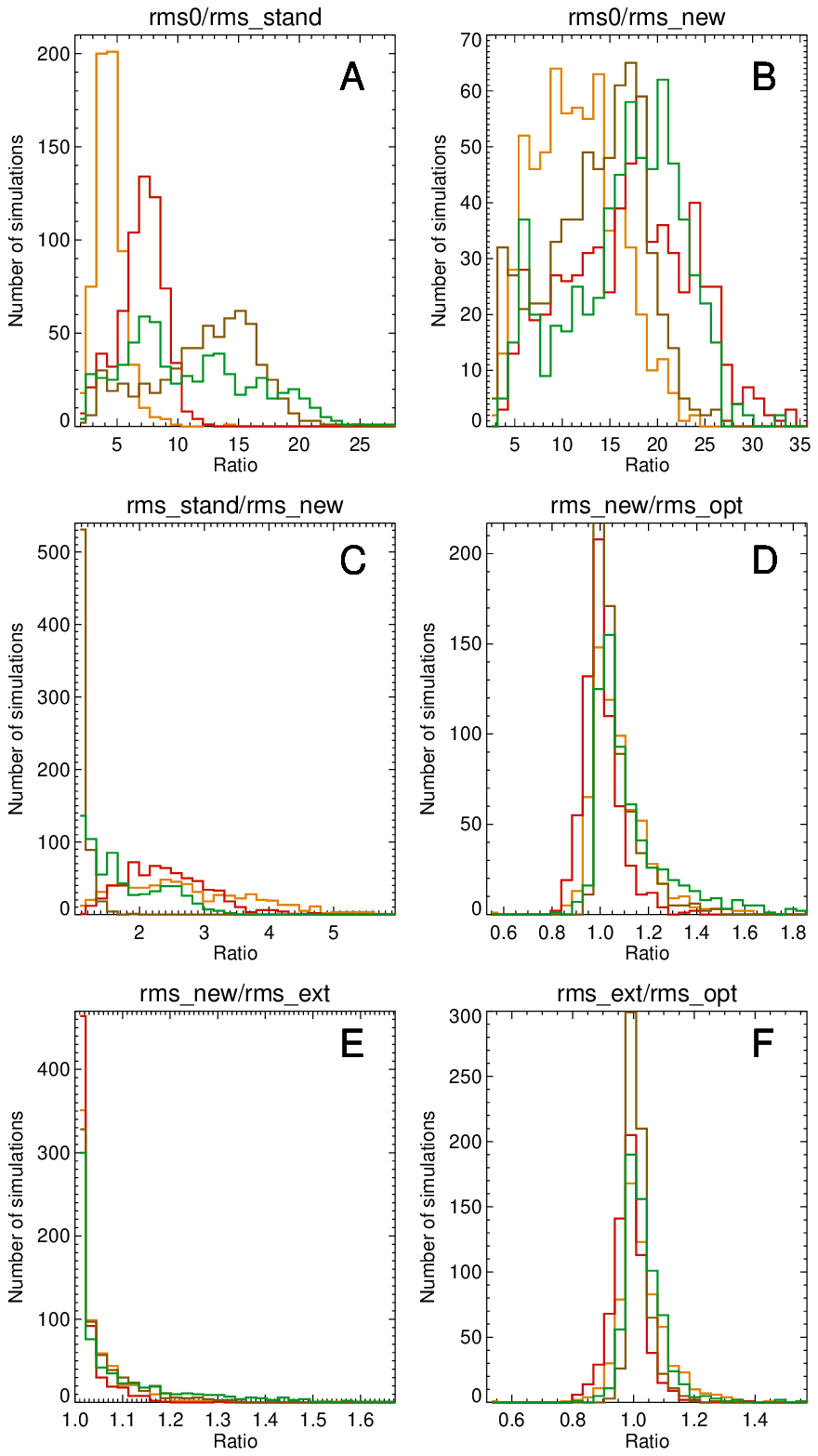}
\caption{
Distribution of different rms ratios, for inclinations of 0$^{\circ}$ (orange), 30$^{\circ}$ (red), 60$^{\circ}$ (brown), and 90$^{\circ}$ (green). 
{\it Panel A:} rms$_0$/rms$_{\rm stand}$.
{\it Panel B:} rms$_0$/rms$_{\rm new}$. 
{\it Panel C:}  rms$_{\rm stand}$/rms$_{\rm new}$.
{\it Panel D:}  rms$_{\rm new}$/rms$_{\rm opt}$.
{\it Panel E:}  rms$_{\rm new}$/rms$_{\rm ext}$.
{\it Panel F:}  rms$_{\rm ext}$/rms$_{\rm opt}$.
}
\label{test_G2}
\end{figure}

\begin{table*}
\caption{Performance for G2, $\Delta$Tspot$_1$ stars}
\label{tab_rms}
\begin{center}
\renewcommand{\footnoterule}{}  
\begin{tabular}{lllllll}
\hline
Inclination  & rms$_0$/rms$_{\rm stand}$   & rms$_0$/rms$_{\rm new}$  &  rms$_{\rm stand}$/rms$_{\rm new}$  &  rms$_{\rm new}$/rms$_{\rm opt}$  & rms$_{\rm new}$/rms$_{\rm ext}$ &   rms$_{\rm ext}$/rms$_{\rm opt}$ \\ 
($^{\circ}$) &    &    &    &    &    &       \\ \hline
  0          &  4.26  & 11.23   &  2.53  & 1.05   &  1.02  &   1.02   \\ \hline
  30         &  7.19  & 17.38   &  2.35  & 1.00   &  1.01  &   0.98    \\ \hline
  60         &  12.82  & 14.13   & 1.08   & 1.04   & 1.02   &  1.01    \\ \hline
  90         &  10.08  & 17.60   & 1.54   & 1.07   & 1.03   &  1.02    \\ \hline
\hline
\end{tabular}
\end{center}
        \tablefoot{Median of the gains between the different correction levels for four inclinations between pole-on (0$^{\circ}$) and edge-on (90$^{\circ}$). These correspond to the distributions shown in Fig.~\ref{test_G2}.  
}
\end{table*}


To illustrate the performance of the new method, we applied this procedure for a subset of G2 star simulations, over one cycle only and $\Delta$Tspot$_1$. We compare the 
rms before correction (rms$_0$), after a standard linear correction with chromospheric emission 
(rms$_{\rm stand}$), and after this new correction method (rms$_{\rm new}$). 
Median values of the gain are shown in Table~\ref{tab_rms} and distributions in Fig.~\ref{test_G2}.
With the new method, the gain with respect to the standard correction is about 2.5 for pole-on configurations and 1.5 for edge-on configurations (median values), and very close to one for inclinations of 60$^{\circ}$ as shown in panel C. This is expected because the reversal of the hysteresis occurs around this inclination, meaning that no improvement is expected in that case. 
The distributions show that for some simulations the gain is as high as 4-6 with respect to the standard correction. 



\subsection{Going beyond the hysteresis correction}

The residuals show that no hysteresis remains, but a curvature is still present (Fig.~\ref{catal}). This is likely to be related to the non-linearity observed in Fig.~\ref{latcst}. After correction of this curvature using a second-degree polynomial in $\log R'_{HK}$  on the residuals shown in Fig.~\ref{catal}, the residuals have a rms of 0.219 and 0.174 m/s (for pole-on and edge-on respectively): this is very close to the residuals after a similar correction (second degree polynomial) for the time series at constant latitude shown in  Fig.~\ref{latcst} (rms of the residuals of 0.215 and 0.176 m/s). We propose to add another step to the procedure, consisting in performing this second degree in the $\log R'_{HK}$  fit, which leads to an rms of the residuals, rms$_{\rm ext}$. The results are shown in Fig.~\ref{test_G2} (panels E and F) and Table~\ref{tab_rms}: for many simulations, the gain is centred on one, but it allows  a significant gain to be added for some of them (tail in panel E).  

After this step, the median gain in power for periods in the habitable zone of G2 stars computed as in \cite{meunier19b}, that is in the 274-777 days range, is between 1.3 (60$^{\circ}$) and 2.8 (30$^{\circ}$). The gain is higher than two (representing a gain of four in mass) for 61, 63, 21, and 49\% of the simulations for the four inclinations respectively. We note that for 7-18\% of the simulations, the power is slightly increased after correction however (the gain is very high for periods between typically P$_{\rm cyc}$/2 and P$_{\rm cyc}$, but not as high below P$_{\rm cyc}$).

As a complementary approach, we also fitted the time series with a function of the form 
RV=$\alpha$(1+$\beta$$\log R'_{HK}$+$\gamma$($\log R'_{HK}$)$^2$)(1+$\delta$t+$\epsilon$t$^2$) where the polynomial in $\log R'_{HK}$  takes a non-linear relationship between RV and chromospheric emission into account, and the polynomial in time plays the role of the reference catalogue functions computed above to correct for the hysteresis. This represents an estimate of the best  gain which can be achieved when considering processes on long (cycle) time scales, as it is less physically constrained than the current functions in the catalogue.  We computed the rms of the residuals after such a correction, rms$_{\rm opt}$. The distribution of the gains are shown in Fig.~\ref{test_G2} (panels D and F) and the median gains in Table~\ref{tab_rms}. The new correction (previous section) provides a gain with respect to  rms$_{\rm opt}$ with a peak around 1, and a small tail toward higher values. After the polynomial fit (rms$_{\rm ext}$), the tail has disappeared, which shows that the correction is very close to what we can expect to reach, which is an excellent result.

We note that at this stage, the rms of the binned series after correction is in the range 0.02-0.33~m/s. There is therefore some signal left, which is more stochastic in nature than what we have removed so far, and is on typically lower timescales. We will study these residuals in a future work. They could be due to departure from the average latitude of large active regions, and/or to size effects (see Fig.~\ref{latcst} and Sect.~6.1). 

In conclusion,  this section shows the feasibility and interest of 
this method, which appears to be very promising and allows substantial improvement of the 
residuals. In a future work, we will test the impact of a degraded sampling (lower number of points, 
incomplete coverage of the cycle). The construction of the catalogue will also be improved using  
smaller steps and/or interpolating in the catalogue, and taking different values of $\theta_{\rm min}$ into account. Finally, the performance 
will also be tested when a planet in the habitable zone is added, and for all spectral types in our 
simulations (F6-K4).

\section{Conclusion}

We analysed a large number of simulated time series of stellar activity covering the spectral types F6 to K4 and different activity levels. We showed that such simulations, based on realistic complex activity patterns, are very useful to understand the limitation of correction methods. We have focused on the methods using the correlation between RV and $\log R'_{HK}$.

The detailed study of the correlation between RV and $\log R'_{HK}$, and the gains in RV jitter which can be obtained using the typical $\log R'_{HK}$  correction compared to the gain obtained when also considering a second-degree polynomial in time, led us to several important conclusions:
\begin{itemize}
\item{Inclination usually plays a crucial role.}
\item{The global and short-term correlations related to the relative weight between  {\it rvspot} and {\it rvplage} (contribution of spots and plages due to their intensity contrast) on one side, and {\it rvconv} (due to the inhibition of the convective blueshift in plages) on the other side, present a complex relationship. This relation is impacted by inclination, and by the addition of other contributions such as oscillation, granulation, and supergranulation.}
\item{Not only is the correction using $\log R'_{HK}$   limited \cite[as shown in previous papers][]{meunier13,dumusque17,meunier19b}, but the addition of a polynomial in time may also help to reduce the RV jitter a little, although in an uncontrolled way. However, the results  show that adding polynomial fits in time should be used with caution.}
\item{An hysteresis between RV and $\log R'_{HK}$  was discovered due to a different relationship between RV and $\log R'_{HK}$  depending on the cycle phase, produced by the combination of geometrical effects (due to the butterfly diagram and the inclination) and the activity level variability along the cycle.  }
\end{itemize}

The  existence of the hysteresis  is a major result, because it comes as a limitation to the RV correction. We have shown that this hysteresis pattern is present in RV solar time series, as well as in other stars. Also, this result will help to better model the long-term RV due to activity. 
 
We propose a new method which significantly improves the correction for long timescales 
(fraction of the cycle, i.e. suitable for planets in the habitable zone). 
The method is based on the physical description of the processes at the origin of the hysteresis (butterfly diagram related to dynamo processes and projection effects), which allows us to better constrain the relationship between the observables (RV and $\log R'_{HK}$) and therefore to better control the residuals, that is, to avoid adding spurious effects due to unphysical contributions.
We note that \cite{haywood16} found that the unsigned magnetic flux (which is an observable  that is currently very difficult to measure on solar-like stars other than the Sun) may be better correlated with long-term RV variation than $\log R'_{HK}$: the difference may be due to a similar effect, with the unsigned flux corresponding to different projection effects. 
The new method proposed in the present paper needs to be further tested on all simulations to derive
realistic detection rates for planets in the habitable zone around F6-K4 stars.  After improvement and validation, these functions can then be made publicly available.  

In addition to these new perspectives to detect low-mass planets in the habitable zone, we will also investigate the possibility of using this method to derive information on the spatio-temporal distributions of stellar activity, in particular in the direction (poleward or equatorward) of the dynamo wave, which  would be extremely interesting since this has been very poorly constrained from observations so far. This should be feasible especially if the stellar inclination is well constrained.

 
Finally, the correction of the RV signal due to activity will be critical for the TESS and PLATO follow-ups.  When  these follow-ups aim at confirming planets observed by transit, this will concern stars seen close to edge-on statistically (unless they have a highly inclined orbit with respect to the equatorial plane of the star) and therefore with a significant hysteresis effect.  For this reason, it is necessary to improve correction techniques taking this effect into account.  



\begin{acknowledgements}

This work has been funded by the ANR GIPSE ANR-14-CE33-0018.
We are very grateful to Charlotte Norris who has provided us the plage contrasts used in this work prior the publication of her thesis. 
The Ca K index was provided by the Sacramento Peak Observatory of the U.S. Air Force Phillips Laboratory(http://nsosp.nso.edu/cak\_mon). 
This work was supported by the "Programme National de Physique Stellaire" (PNPS) of CNRS/INSU co-funded by CEA and CNES.
This work was supported by the Programme National de Plan\'etologie (PNP) of CNRS/INSU, co-funded by CNES.
This research has made use of the SIMBAD database, operated at CDS, Strasbourg, France.

\end{acknowledgements}

\bibliographystyle{aa}
\bibliography{biblio}

\begin{appendix}

\section{RV--$\log R'_{HK}$  correlations}

        This Appendix provides further details of the RV--$\log R'_{HK}$ correlations for our simulations: the correlation is computed either on whole time series (global correlations) or on subsets of data (local correlations), as discussed in Sect.~3. We then quantify these correlations for the Sun.

\subsection{Impact of the parameters on the global correlation}

\begin{figure}
\includegraphics{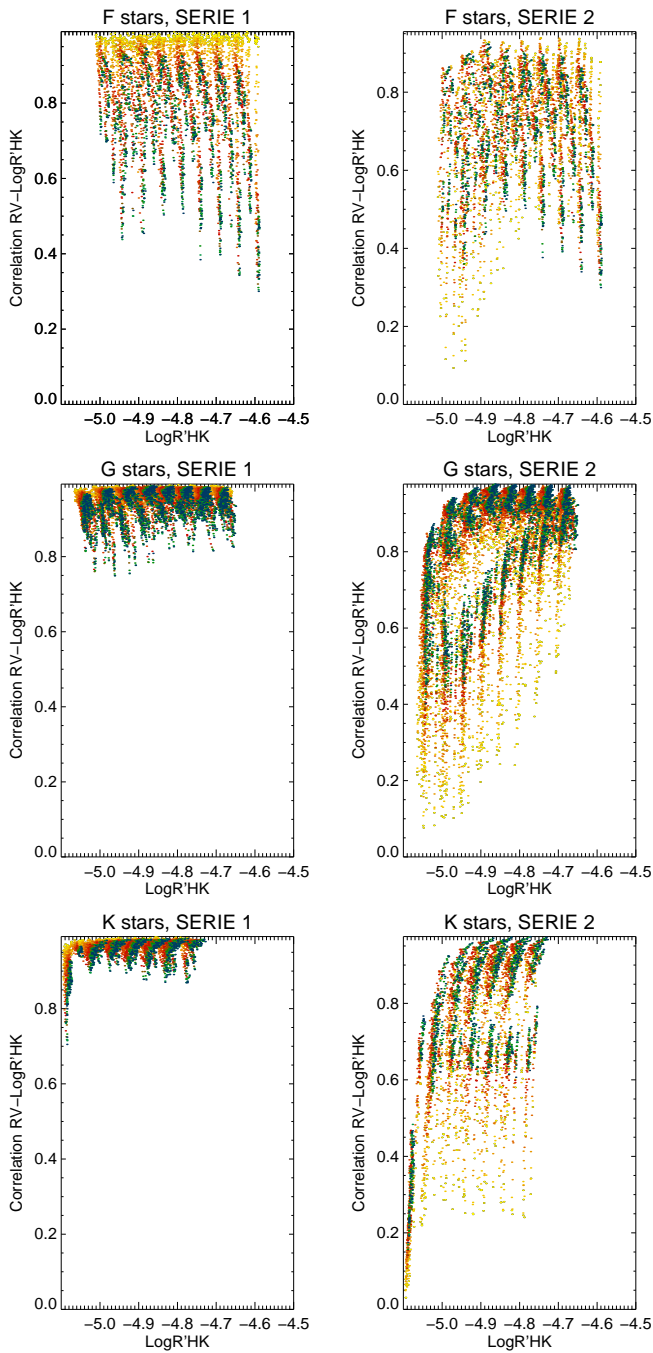}
\caption{
Correlation between RV (due to activity only, SERIE 1, left, and including {\it rvogs} and {\it rvinst}, SERIE 2, right) and $\log R'_{HK}$  vs. average $\log R'_{HK}$  for different types of stars: F stars (upper panels), G stars (middle panels), and K stars (lower panels). 
        The colour code  is similar to that of Fig.~\ref{gain_rv}.
         Only one point out of five is shown for clarity. 
}
\label{correl}
\end{figure}

The left panels of Fig.~\ref{correl} show the global  correlation for RV due to activity only ({\it rvspot2}, corresponding to $\Delta$Tspot$_2$ as defined in Sect.~2.1), computed over each time series, as a function of the average activity level. The plots for $\Delta$Tspot$_1$ are very similar (not shown here). The correlation is usually high (above 0.7), but there is a drop in correlation for active F stars and high inclinations. This is due to the fact that for these stars the spot and plage contribution at the rotation period (which is not correlated with $\log R'_{HK}$) increases relative to the convection inhibition contribution: this decreases the correlation.

After adding the OGS signal as well as instrumental white noise (right panels), we observe a decrease of the correlation values for both quiet and active stars of all spectral types. The effect is particularly strong for K stars, despite the fact that the OGS signal is also decreasing when going to lower mass stars (see Paper I). 

\subsection{Relationship between correlation and convective RV component}

\begin{figure}
\includegraphics{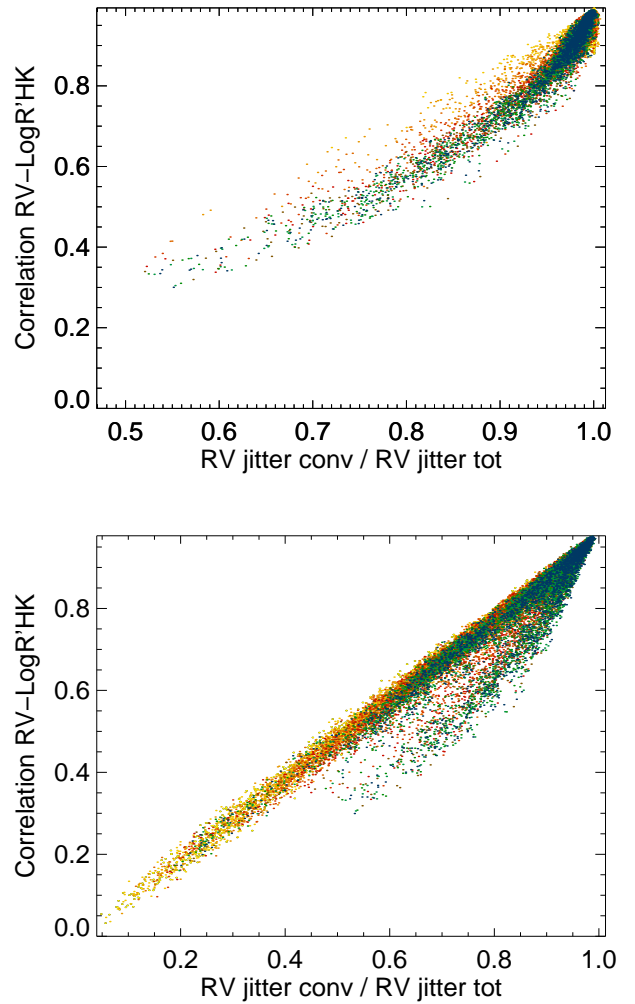}
\caption{
{\it Upper panel:}
        Correlation between RV (due to activity only) and $\log R'_{HK}$ vs. RV jitter due to convection only ({\it rvconv}) divided by total RV jitter (due to activity only).  The colour code is similar to  Fig.~\ref{gain_rv}.
{\it Lower panel:} Same including {\it rvogs} and {\it rvinst} (for both the total RV jitter and the RV used to compute the correlation).
 Only one point out of five is shown for clarity. 
}
\label{correl_rv}
\end{figure}

Figure~\ref{correl_rv} shows the correlation between RV and $\log R'_{HK}$  vs. the ratio between the RV jitter due to convection alone (which is the component which should be correlated to $\log R'_{HK}$) and the total RV jitter. With no-noise added (i.e. no OGS nor instrumental noise), there is a relatively good correlation between the two (0.92), but the relation is not entirely linear. The contribution of {\it rvconv} can be as low as 40-50\% over our range of parameters, and the correlation between RV and $\log R'_{HK}$, although always positive, can reach values of 0.2 (and even almost 0 in the presence of noise). 

In the presence of noise (lower panel), the relationship is more complex. The global correlation remains good for two thirds of the simulations (66\% still have a correlation above 0.7, while this was 95\% for the no-noise case), but both ratios and correlations reach lower values (down to 0). Moreover, we see two regimes: one with a very good correlation and a linear relationship, and one below, where the correlation is degraded. 
This lower regime corresponds to F stars (37\% of the F stars are in the second regime) with high inclinations (i.e. close to edge-on) compared to average. This is due to the fact that the correlation seems to be more strongly affected by inclination than the RV jitter ratio for these stars, leading to the observed shift toward lower correlations.
This is probably due to the fact that in some cases, the addition of the OGS signal significantly degrades the correlation even though its contribution to the RV jitter is not major.

\subsection{Short-term correlation}

\begin{figure}
\includegraphics{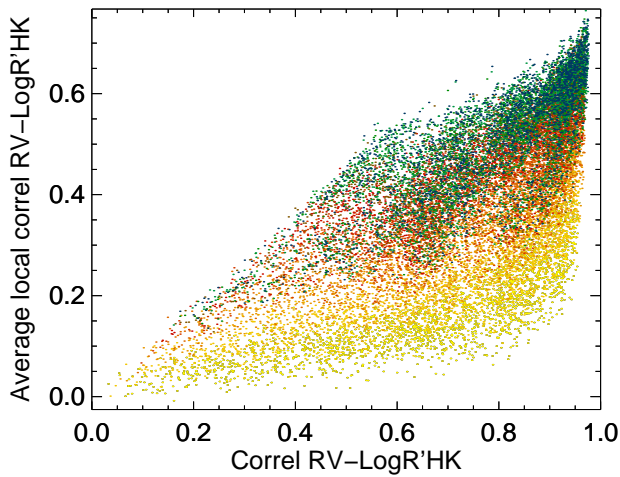}
\caption{
Average short-term correlation between RV (including activity, {\it rvogs} and {\it rvinst}) and $\log R'_{HK}$ vs. global correlation. 
        The colour code is similar to  Fig.~\ref{gain_rv}.
        Only one point out of five is shown for clarity. 
}
\label{correl_loc}
\end{figure}

\begin{figure}
\includegraphics{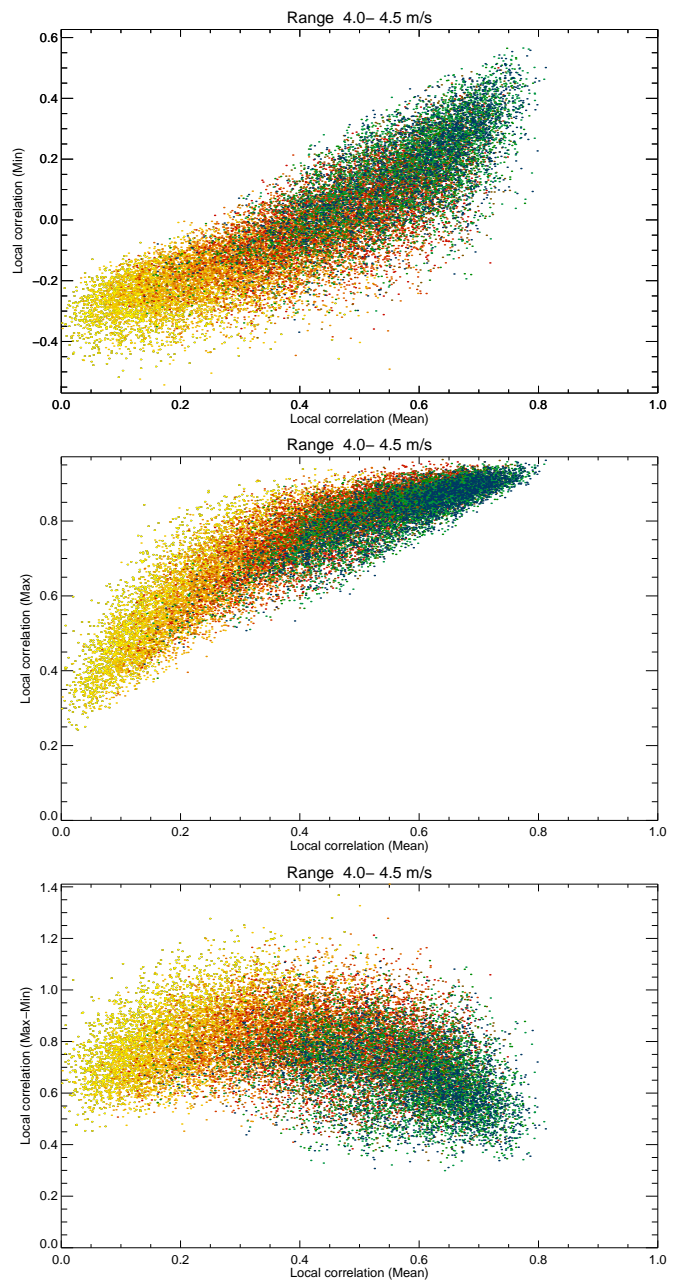}
\caption{
        {\it Upper panel:} minimum value of short-term correlation (including activity, {\it rvogs} and {\it rvinst}). The minimum is computed over the $N$ values of a given time series (see text). The colour code is similar to  Fig.~\ref{gain_rv}.
{\it Upper panel:} Same for the maximum of the short-term correlations.
{\it Lower panel:} Same for the maximum minus minimum of the short-term correlation.
         Only one point out of five is shown for clarity. 
}
\label{var_correl}
\end{figure}

We have so far considered the global correlation between RV and $\log R'_{HK}$  . This global correlation includes contributions from both short and long timescales. We now compute a short-term correlation, which is sensitive to the short (P$_{\rm rot}$) timescales but not to the longer timescales: it is impacted by the convective blueshift inhibition in plages (correlated with $\log R'_{HK}$) and by the spot+plage signal, as well as the OGS signal when considered (not correlated with $\log R'_{HK}$). We compute the short-term correlation over time spans of 90 days: this is repeated for $N$ consecutive intervals over the whole time series, with $N$ between 35 and 59 depending on the length of time series. The average of the $N$ values over each time series is then computed. The resulting averaged short-term correlation is shown in Fig.~\ref{correl_loc}.


The short-term correlation is related to the global correlation, however there  is a large number of simulations with short-term correlations much lower than the global one. 
We also note that if the inclination of a given star is well known, it is possible to use this plot to relate the global and short-term  correlations. 

For RV including the OGS signal and the instrumental noise, the short-term correlations can vary within a range of 0.4 to 1 depending on the time series, as shown in Fig.~\ref{var_correl}. For example, for an average short-term correlation of 0.6, the minimum varies between -0.2 and 0.4, and the maximum between 0.75 and 0.95. For an average short-term correlation, the minimum is typically in the range -0.4 to -0.1 (and therefore negative) and the maximum between 0.45 and 0.8. 

\subsection{The solar case}

\begin{figure}
\includegraphics{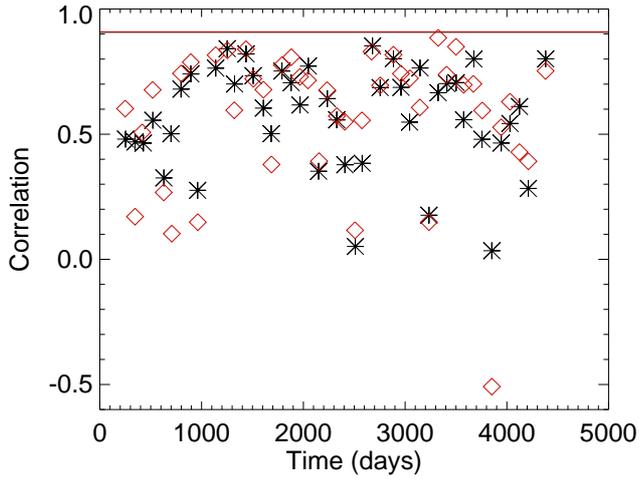}
\caption{
Local correlation between RV and $\log R'_{HK}$  vs. time (Julian days - 2450000) for RV reconstructed from observed solar structures \cite[][]{meunier10a} in black (stars) and from RV estimated from MDI/SOHO Dopplergrams \cite[][]{meunier10} in red (diamonds). The horizontal lines correspond to the global correlations. 
}
\label{correl_sol}
\end{figure}

\begin{figure}
\includegraphics{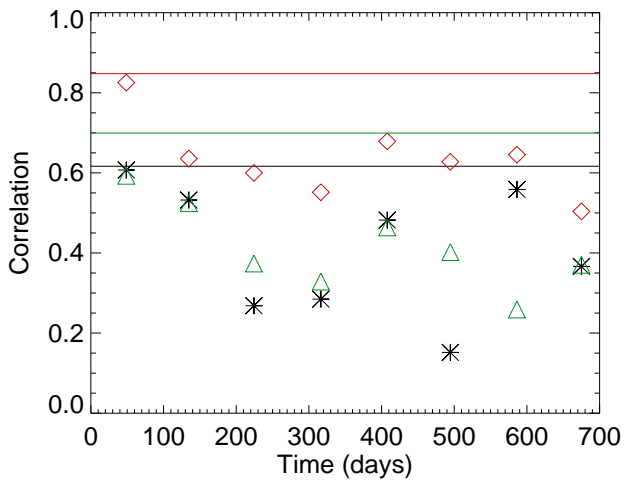}
\caption{
Local correlation between RV and $\log R'_{HK}$  vs. time (Julian days - 2457222.5) for RV observed with HARPS (black stars), and reconstruted from HMI (raw in red diamonds, with noise added in green triangles), estimated from time series published by  \cite{milbourne19}.  The horizontal lines correspond to the global correlations. 
}
\label{correl_milbourne}
\end{figure}

Figure~\ref{correl_sol} shows the local correlation in the solar case for the two time series obtained by \cite{meunier10a} and \cite{meunier10} covering a complete cycle. They are  described in Sect.~5.3. The average local correlation is similar (0.57-0.58) while the global correlation is much higher (0.91),  which agrees  with our simulations (see Fig.~\ref{correl} and \ref{correl_loc}). The dispersion of the correlation values is respectively 0.20 and 0.27 and the local correlations cover a wide range of values as in our simulations. 

Similar local correlations can be computed from the HARPS solar observation published by \cite{milbourne19}, and are shown in Fig.~\ref{correl_milbourne}; the average of around 0.41 is slightly lower than for the previous time series. The correlation with the HMI/SDO RV that these latter authors reconstructed (no noise, spot, plage and convective blueshift inhibition only) is larger: on average around 0.63. However, if a noise of 1.2 m/s is added for example, the average local correlation decreases (0.39, similar to the correlation with HARPS). The local correlations also show a wide dispersion, as in our previous results.


\end{appendix}

\end{document}